\documentclass{basi}
\usepackage{epsfig}

\begin{document}
\title[Comparison of Energies]{Comparison of Energies Between Eruptive Phenomena and Magnetic Field in AR 10930} 
\author[Ravindra and Howard]%
       {B. Ravindra\thanks{e-mail:ravindra@iiap.res.in} and Timothy A. Howard$^{1}$ \\ 
        Indian Institute of Astrophysics, Koramangala, Bangalore 560 034 \\ 
        $^{1}$Department of Space Studies, Southwest Research Institute, Boulder, CO 
80302, USA \\}


\maketitle
\label{firstpage}
\begin{abstract}
We present a study comparing the energy carried away by a coronal mass ejection 
(CME) and the radiative energy loss in associated flare plasma, with the decrease 
in magnetic free energy during a release in active region NOAA 10930 on December 13, 2006 during the declining phase of the solar cycle 23. The ejected CME
was fast and directed towards the Earth with a projected speed of $\sim$1780~km~s$^{-1}$ and a de-projected speed of $\sim$3060~km~s$^{-1}$. We regard these as lower and upper limits for our calculations. It was accompanied by an X3.4~class flare in the active region. The CME carried (1.2--4.5)$\times$10$^{32}$~erg (projected-deprojected) of 
kinetic and gravitational potential energy. The estimated radiative energy loss during 
the flare was found to be 9.04$\times$10$^{30}$~erg. The sum of these energies 
was compared with the decrease in measured free magnetic energy during the flare/CME. The free energy is that above the minimum energy configuration and  was estimated using the 
magnetic virial theorem. The estimated decrease in magnetic free energy is large,
3.11$\times$10$^{32}$~erg after the flare/CME compared to the pre-flare energy. Given the range of possible energies we estimate that 50--100\%\ of the CME energy arose from the active region.
The rest of the free magnetic energy was distributed among the 
radiative energy loss, particle acceleration, plasma and magnetic field reorientation.
\end{abstract}

\section{Introduction}
\label{sec:intro}
Sunspots are the harbor for solar magnetic fields. Energetic phenomena like
solar flares occurring in sunspots derive their energy from the active region 
magnetic fields. It is generally believed that coronal mass ejections (CME) which are
associated with active regions obtain the majority, if not all of, their energy
from these magnetic structures. The magnetic energy stored in the twisted 
magnetic fields is released within a few minutes during these energetic events and
is of the order of 10$^{32}$~erg. The majority of the energy released is carried 
away by the CME mostly in the form of kinetic energy and partly as gravitational potential energy. The rest of the energy is mostly utilized in particle acceleration, plasma 
heating, radiation and magnetic field re-orientation (Webb et~al. 1980, Canfield et~al. 
1980, Emslie et al. 2004). 

Most of the energetic processes occur in the corona. To understand
these processes it is essential to measure the magnetic energy and its temporal
evolution. However, measurement of the coronal magnetic field is difficult because
of observational constraints (Solanki et~al. 2003, Lin et~al. 2004). 
Currently, only the magnetic field measurements 
are accurately obtained at the photospheric levels with sufficient resolution and
cadence. With the available magnetograms and using many techniques it is, however, possible 
to estimate the available magnetic energy in the corona with some uncertainties.

There are many methods to estimate the available magnetic energy in the active 
region corona. These are: (1) the minimum-current corona model (Longcope 
et~al. 2007, Kazachenko et~al. 2009); (2) the non-linear force free field (NLFFF)
extrapolation and using the volume integral (Srivastava et~al. 2009); (3) the Poynting flux 
estimation (Ravindra, Longcope and Abbett 2008); and (4) the magnetic virial 
theorem (Sakurai 1987, Metcalf et~al. 1995, Metcalf, Leka and Mickey, 2005, 
Venkatakrishnan and Ravindra, 2003). Among these, the energy estimation 
using the  minimum-current corona model requires only the line-of-sight magnetic 
field measurements and it provides only the energy involved in the flare. All other 
techniques require the vector magnetic field measurements taken at the photosphere 
or chromosphere.

The energy estimation using the Poynting flux requires not only 
the magnetic vector field measurement, but also the magnetic footpoint 
velocity. The remaining two techniques rely on vector magnetograms obtained 
at one level, preferably in the chromosphere where the magnetic field is 
force-free. Up until very recently, the vector magnetic field measurements have been 
routinely made at the photospheric level. With the 
advent of better computing facilities and readily available photospheric data, 
researchers have developed algorithms and applied them to photospheric vector 
magnetograms to mimic the chromospheric data (Wiegelmann, Inhester and Sakurai 
2006). The procedure is known as the ``pre-processing'' of the vector field 
data. This technique was further improved when the chromospheric H$_\alpha$ 
data became available (Wiegelmann et~al. 2008). The resulting magnetograms 
are used as lower boundary data for the NLFFF extrapolation to compute the 
coronal magnetic fields (Derosa et~al. 2009). After 
computing the coronal magnetic fields it is a relatively simple task to estimate 
the free energy available within the volume (Schrijver et~al. 2008). 

The magnetic virial theorem can be applied to estimate the magnetic energy provided 
the field measurements are made at a level where the magnetic force and
torque vanish (Molodenskii 1969, Sakurai 1987, Metcalf et~al. 1995). In the 
past, many researchers have used the magnetic virial theorem to compute the free 
magnetic energy (Gary et~al. 1987, Metcalf et~al. 1995, Metcalf, Leka and Mickey 
2005). Metcalf et~al. (1995) and Metcalf, Leka and Mickey (2005) used the 
chromospheric vector magnetic field data observed in the Na~I absorption line
and these are considered to be close to the force-free 
condition (Metcalf et~al. 1995). Using the chromospheric vector magnetograms they 
not only verified the force-free conditions, but they also provided a measurement of 
the magnetic 
energy. The magnetic free energy was measured in NOAA AR 10486 (a super active region) 
on 29 October 2003 before and after the flare (Metcalf, Leka and Mickey 2005). They 
found an increase in magnetic free energy after the flare compared to the pre-flare 
energy level. 

Space weather prediction is a major challenge to the solar community as the major
eruptive events on the Sun can produce large geomagnetic storms. Since the eruptive
events derive their energy from the active regions, it is essential to measure the 
available free energy in the active region volume. It is also essential to validate this
based on the energy carried away by the eruptive processes (Emslie et al. 2004). In this
paper, we compare the drop in free magnetic energy with the energy content of the CME and 
thermal plasma during the X3.4~class flare produced by AR 10930. 

\subsection{Active Region 10930}

The active region NOAA 10930 produced several X-class and M-class flares during its passage across the Sun in December 2006. It was well observed by Hinode and SOHO whose data sets 
are not influenced by seeing and 
Earth's diurnal effects. This active region has been extensively studied for 
the measurement of twist (Su, et~al. 2009), 
rotation (Min and Chae 2009), penumbral filaments (Tan et~al. 2009), flares 
(Isobe et~al. 2007) and helicity (Magara and Tsuneta 2008). Space based observations, 
free from seeing induced polarization, better techniques in measuring the Stokes vectors, 
improved sensitivity of the instrument, good spatial resolution, improved inversion 
and ambiguity resolution techniques, and techniques to mimic the vector field measured 
at the photospheric level to closely match the vector field measured at the lower 
chromosphere, all provide good data sets for the measurement of free magnetic energy 
in AR 10930 before and after the X3.4~class flare occurred on Dec 13, between 02:14
and 02:57~UT.   

In this paper, we present the study of the temporal evolution of magnetic energy computed over a 
period of 5 days surrounding the flare and associated Earth-directed CME. We used the 
Hinode/SP vector magnetograms to estimate the
magnetic free energy available in the active region during this period. We applied a 
magnetic virial theorem algorithm to the
pre-processed vector magnetic field measurements. In Section 2, we use vector 
magnetograms computed from Hinode/SP to estimate the amount of accumulated magnetic
free energy during the evolution of AR~10930 from December 9-14, 2006. This time
frame includes the flare and CME occurred on December 13. To better understand the 
decrease in free energy from this 
active region, we estimate the amount of kinetic and gravitational potential energy
carried by the CME using the time series of LASCO images.
In Section 3, we present the evolution of magnetic free energy using the magnetic 
virial theorem over the time period. A comparison of the decrease in magnetic free energy 
after the event with the energy carried away by the CME and the radiative energy 
loss during the flare is performed. In the last Section, we conclude with discussions 
on the energy partition of the CME and flare.

\section{Observational Data and Analysis Methods}
\subsection{Hinode/SOT}
The solar optical telescope (SOT) on board the Hinode satellite (Kosugi et~al. 2007) 
makes spectro-polarimetric measurements at a resolution of 0.3$^{\prime\prime}$ 
(Ichimoto et~al. 2008). The active region maps have been produced by spatial 
scanning and the Stokes I, Q, U and V spectra are obtained in the Fe~I~6301.5~\AA~and 
6302.5~\AA~absorption lines. The spectro-polarimeter is capable of making the raster 
scan of the active region in the fast mode as well as in the normal mode. In the 
fast mode, spatial resolution along the slit direction is 0.295$^{\prime\prime}$ and 
in the scanning direction is 0.317$^{\prime\prime}$/pixel. The fast mode scanning 
takes typically 1~hr and the normal mode takes 2~hr to complete the I, Q, U and V 
measurements over the active region. Our data set consists of only the fast mode data. 
The Stokes data were obtained from 9--14 December 2006. We chose the 
period at which the active region was located in the solar longitude range of E30 to W30. 
During this period 24 magnetograms were obtained from the fast mode. The
Stokes signals were calibrated using the standard solar software pipeline for 
the spectro-polarimetry. In order to get the complete information on the 
vector magnetic fields, the obtained Stokes vectors have been inverted using the
Milne-Eddington inversion (Skumanich and Lites 1987, Lites and Skumanich 
1990, Lites et~al. 1993). The 180$^{\circ}$ ambiguity was resolved based on the 
minimum energy algorithm (Metcalf 1994) implemented by Leka, Barnes and Crouch 
(2009). The resulting magnetic field vectors have been transformed to heliographic coordinates as described by Venkatakrishnan, Hagyard and Hathaway (1988). 
The resulting B$_z$ (line-of-sight field) and transverse field strength has 
1-$\sigma$ error bars of 8~G and 30~G respectively. These vector field data have been 
used in computing the available magnetic energy.

SOT/SP makes the vector magnetic field measurement at the photospheric level where
the force-free assumption is invalid (Metcalf et~al. 1995). Since the measurements of the
vector magnetic fields at the chromospheric/coronal level are not available routinely, 
we have used the field measurements made at the photospheric
level. Wiegelmann, Inhester and Sakurai (2006) have developed a method 
of pre-processing to mimic the photospheric vector magnetic field data so that they 
appear to be force-free. We have applied the pre-processing method described in 
Metcalf et~al. (2008) to the vector magnetograms to minimize the effect of the 
magnetic force and torque in the data. The procedure minimizes the 2-D functional as in 
their Equation (7). This function consists of 4 terms and each constraint is weighted by 
a undetermined factor. The first and second term in their equation corresponds to force 
balance and the torque-free condition. The last two terms correspond to optimized 
boundary condition and smoothing parameters. The resulting pre-processed data set is close 
to force free and torque free (see Section~3.1) but not necessarily in each and every point 
in the magnetogram. 

In order to estimate the magnetic energy for the force-free media we have used 
the magnetic virial theorem (Chandrasekhar 1961, Molodenskii 1969, Aly 1984 and 
Low 1985), given by

\begin{equation}
E = \frac{1}{4\pi}\int(xB_{x}+yB_{y})B_{z} \mbox{d}x\mbox{d}y,
\end{equation}

where, $E$ is the magnetic energy, $B_{x}$ and $B_{y}$ are components of the 
horizontal magnetic fields and $B_{z}$ is the vertical component of the field. 
The free magnetic energy is the excess energy above the minimum energy state 
(energy content of the potential field). The free energy $E_{free}$ can 
be estimated by taking the difference in the total available energy and 
the magnetic potential energy as

\begin{equation}
E_{free} = E_{Tot} - E_{Pot}
\end{equation}

where $E_{Tot}$ is the total accumulated magnetic energy and $E_{Pot}$ 
is the magnetic potential energy. While computing the energy using the virial theorem we
opted for the center of FOV of the magnetogram as 
the origin. The potential magnetic fields have been derived from the $B_{z}$ component by
using the Fourier transform method (Alissandrakis, 1981). The routine $\it fff.pro$ 
available in the solar software pipeline has been adopted here to compute the potential
component of the magnetic field, $B_{xp}$ and $B_{yp}$. This routine is written based
on the constant $\alpha$ force-free field method of Alissandrakis (1981) and Gary (1989).
The pre-processed $B_{z}$ component is used in the computation of $B_{xp}$ and $B_{yp}$.
Using $B_{z}$, $B_{xp}$, 
$B_{yp}$ and Equation~(1) we have computed the energy of the magnetic potential field. For the
computation of total magnetic energy we have used the pre-processed vector magnetograms.

\subsection{SOHO/LASCO Data}
To analyze the associated coronal mass ejection we have utilized space-based 
coronagraphs. The Large Angle Spectroscopic Coronagraph (LASCO) on board SOHO 
(Brueckner et~al. 1995) currently consists of two coronagraphs that block out 
the majority of the bright photospheric light from the Sun to reveal the surrounding corona. 
The C2 coronagraph has a field of view of 2.0--6.0 R$_{\odot}$ and a cadence of 
approximately 30 minutes, while C3 has a field of view of 3.7--30 R$_{\odot}$ 
and a cadence of $\sim$50 minutes (where, R$_{\odot}$ corresponds to 1-solar radius).
Both are white light telescopes, i.e.\ they
observe the Thomson scattered light from free electrons in the plasma
comprising the corona. Density enhancements in the corona such as coronal mass 
ejections (CMEs) can be identified with relative ease, and the distance from the
Sun and mass of the CME can be estimated. LASCO has shown to be highly 
successful at identifying and tracking CMEs, with well over $10^4$ detected 
to date (see Yashiro et~al. 2004, Howard et~al. 2008 or the CDAW CME catalog at 
http://cdaw.gsfc.nasa.gov/CME\_list).

\begin{figure*}
\begin{center}
\includegraphics[width=60mm]{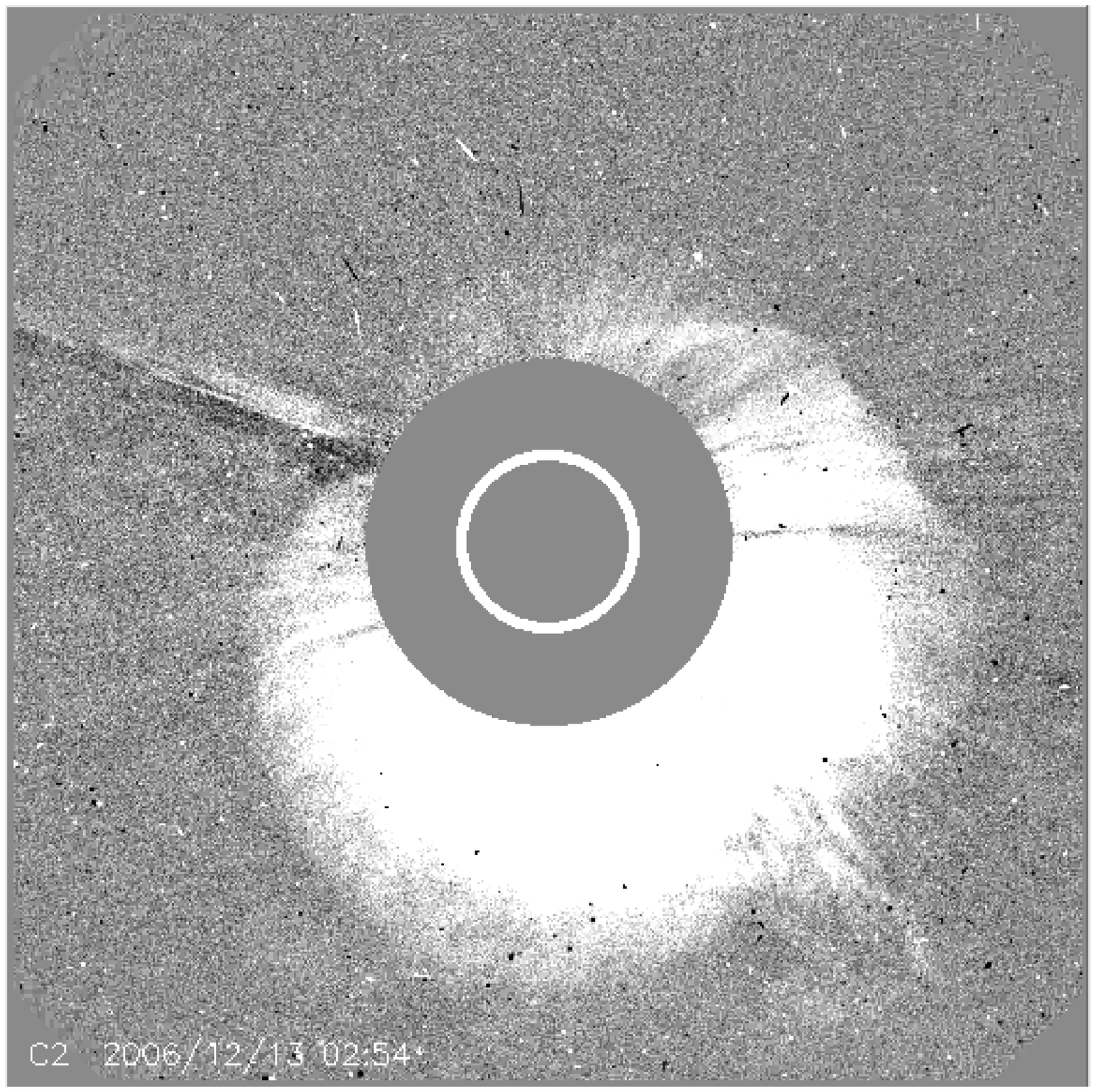}\includegraphics[width=60mm]{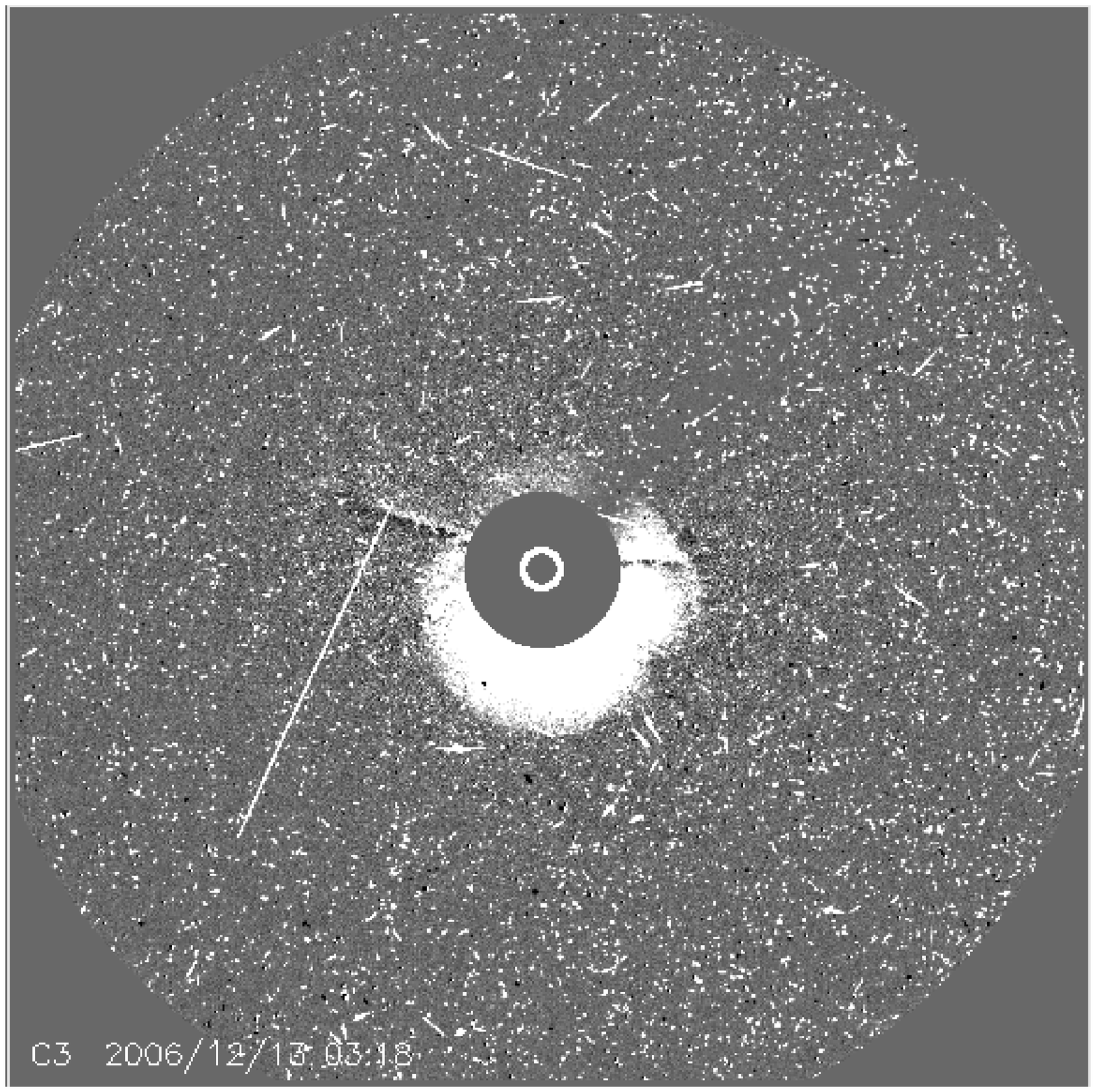} \\
\end{center}
\caption{A halo coronal mass ejection observed by the LASCO C2 coronagraph at 02:50~UT 
(left) and C3 at 03:18~UT (right) on December 13, 2006. These are running difference images
where the previous image has been subtracted from the present. The white circle at the center 
of each 
image represents the solar surface and the solid grey circle represents the occulting disk. 
In C3 the so-called ``snow storm'', due to high-energy particles striking the LASCO detector, is visible. These later saturate the images in C2 and C3 to the point where no measurements are possible.}
\label{fig:1}
\end{figure*} 

By applying well-established assumptions about the observed CME we may 
identify its distance evolution and mass, from which an estimate of its 
kinetic and gravitational potential energy may be produced. Distance measurements are 
obtained by measuring the location of the leading edge of the CME relative 
to the Sun. Most workers choose a fixed direction or single location on the 
CME front at which to make these measurements (e.g. Yashiro et~al. 2004). Mass 
is estimated by measuring the intensity across the entire area enclosing the 
CME and converting it to a density using Thomson scattering theory
(e.g. Billings 1966).

Projection effects play an important role in the observed structure of the CME 
(e.g. Howard and Tappin 2008). LASCO images are projected into the plane of 
the sky, and so the closer the CME is to the Sun-Earth line, the more heavily 
projected its images will be. For mass calculations the angle to the plane of 
the sky is an integral component of the coefficients involved in the Thomson 
scattering physics (angle $\chi$ in Billings (1966)), and so the direction of 
propagation can be easily included in the calculations. For identifying 3-D 
distance properties we may apply spherical geometry provided we have an idea 
of the direction of propagation of the CME. For the analysis in the present 
study we apply the technique of Howard et~al. (2007, 2008) 
who used the location of the associated solar surface eruption as the direction 
of propagation. In this case, this is the flare in AR 10930, which was located 
at 6$^{\circ}$S, 23$^{\circ}$W at the time of the eruption. It is important to note that the location
of the flare is not an excellent indicator of 
CME direction as it has been 
well established that flares are more usually associated with the footpoint of 
a CME (e.g. Harrison 1986). In the absence of other 3-D directional information, 
however, the location of the flare is a reasonable approximator.

Because of these and other uncertainties we consider two values for speed and mass and consequently for the energies: the projected values (unchanged from the original height-time measurements); and de-projected values following the analysis below. These two values may be regarded as lower and upper limits to the calculated values of speed, mass and energy, respectively.

Following Howard et~al. (2007) we use the following equation for 3-D distance:

\begin{equation}
{1\over{R}}=\sin\alpha\cot\varepsilon + \cos\alpha,
\label{1overR}
\end{equation}
where $R$ is the radial distance of the measured point (on the CME) from the 
Sun in AU and $\alpha$ is the angle subtended by the measured point at the Sun, 
given by
\begin{equation}
\cos\alpha = \cos\theta\cos\phi,
\end{equation}
where $\theta$ and $\phi$ are the latitude and longitude of the direction of propagation vector, which 
in this case is the location of the flare site. The angle $\varepsilon$ is the 
elongation, which is easily obtained via the application of two simple assumptions 
about CMEs observed by coronagraphs. These are the Point P approximation 
(e.g. Houminer and Hewish 1972) and 
that $\varepsilon$ is small. This enables the simple conversion 
$r$ (AU) = $\varepsilon$ (rad), where $r$ is the projected distance of the 
measured point (Howard et~al. 2007). To determine the mass we use the standard codes
for mass calculation available via the Solarsoft software suite (including $eltheory.pro$). 
These codes are based
on the theory of Billings (1966) (page 150) and include the $\chi$ term for direction.

The CME associated with the flare on December 13 was Earth-directed event and 
first observed by LASCO C2 at 02:54~UT on December 13. 
Assuming a constant speed its projected onset was 01:49--02:19 ~UT (de-projected--projected), within the 8 hours between the vector magnetic
field measurements. Figure~\ref{fig:1} shows LASCO C2 and C3 images of 
the CME. Shortly after launch, energetic particles from the flare bombarded SOHO resulting 
in the saturation known as a ``snow storm''. The bombardment intensified reaching 
a point where the LASCO cameras were almost completely saturated by around 
04:00~UT. As a result, we were only able to obtain mass measurements from single C2 and C3 images (at 02:54 and 03:18~UT from Figures~\ref{fig:1}
respectively) and could only obtain height (or distance) -- measurement until 04:18~UT. 

\section{Results}

\begin{figure*}
\begin{center}
\includegraphics[width=100mm]{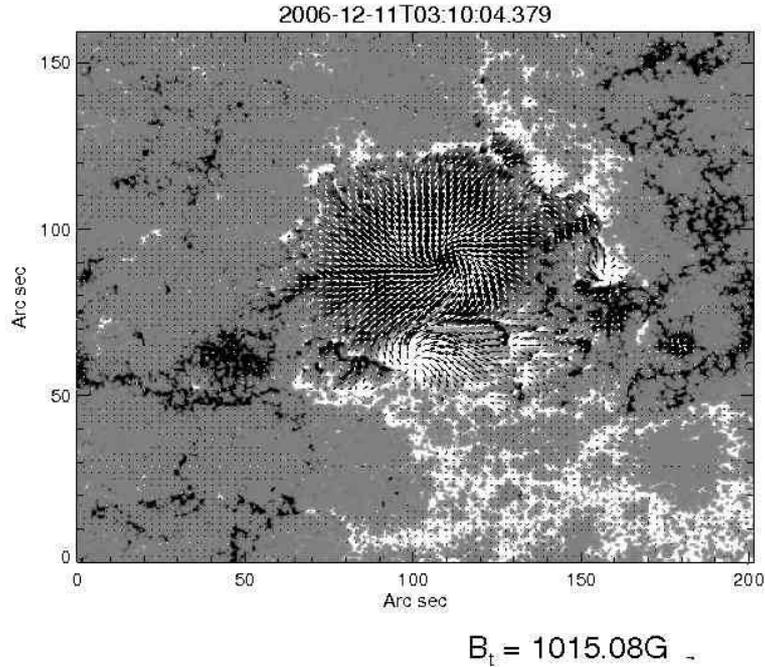}
\end{center}
\caption{A map of the vector magnetic field of the active region 10930. The transverse field vectors are overlaid upon the B$_{z}$ component of the magnetic field.
The black (white) colors represent the south (north) polarity regions.}
\label{fig:2}
\end{figure*}

The active region NOAA 10930 appeared on the Eastern limb of the Sun on 
December 5, 2006 in the southern hemisphere at a latitude of 5$^{\circ}$, with 
two polarities that were almost perpendicularly aligned with the East-West direction. 
It was visible on the solar disk until December 17 when it crossed the West limb of the sun (see: http://www.solarmonitor.org/index.php). Figure \ref{fig:2} 
shows the sample image of the vector magnetogram with the transverse field vectors 
overlaid upon B$_{z}$ for the Dec 11 observation.  The vertical magnetic field strength 
in the South (negative) and North (positive) polarity sunspot is in excess of -4000 
and 2300~G respectively.  

In the vector field map (Figure \ref{fig:2}) it is clear that the field lines are twisted.
Both the polarities are located close to each other and the boundary between the two
is highly sheared. The North polarity region is emerging in the Southern part of the 
existing South polarity region. As it emerges it rotates in the anti-clockwise 
direction and the transverse fields are twisted in the clockwise direction.

\subsection{Magnetic Energy}

\begin{figure*}
\begin{center}
\includegraphics[width=60mm]{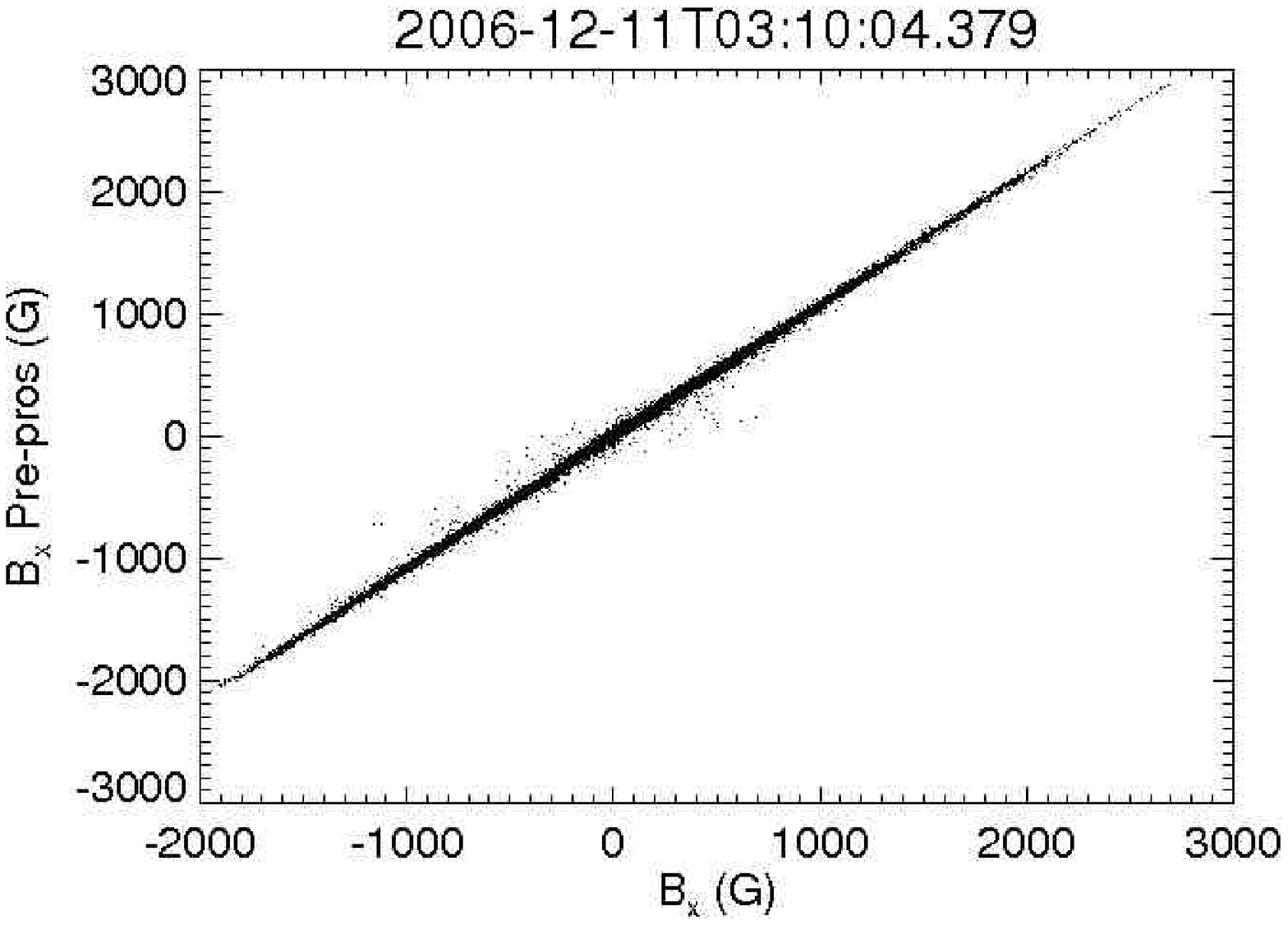}\includegraphics[width=60mm]{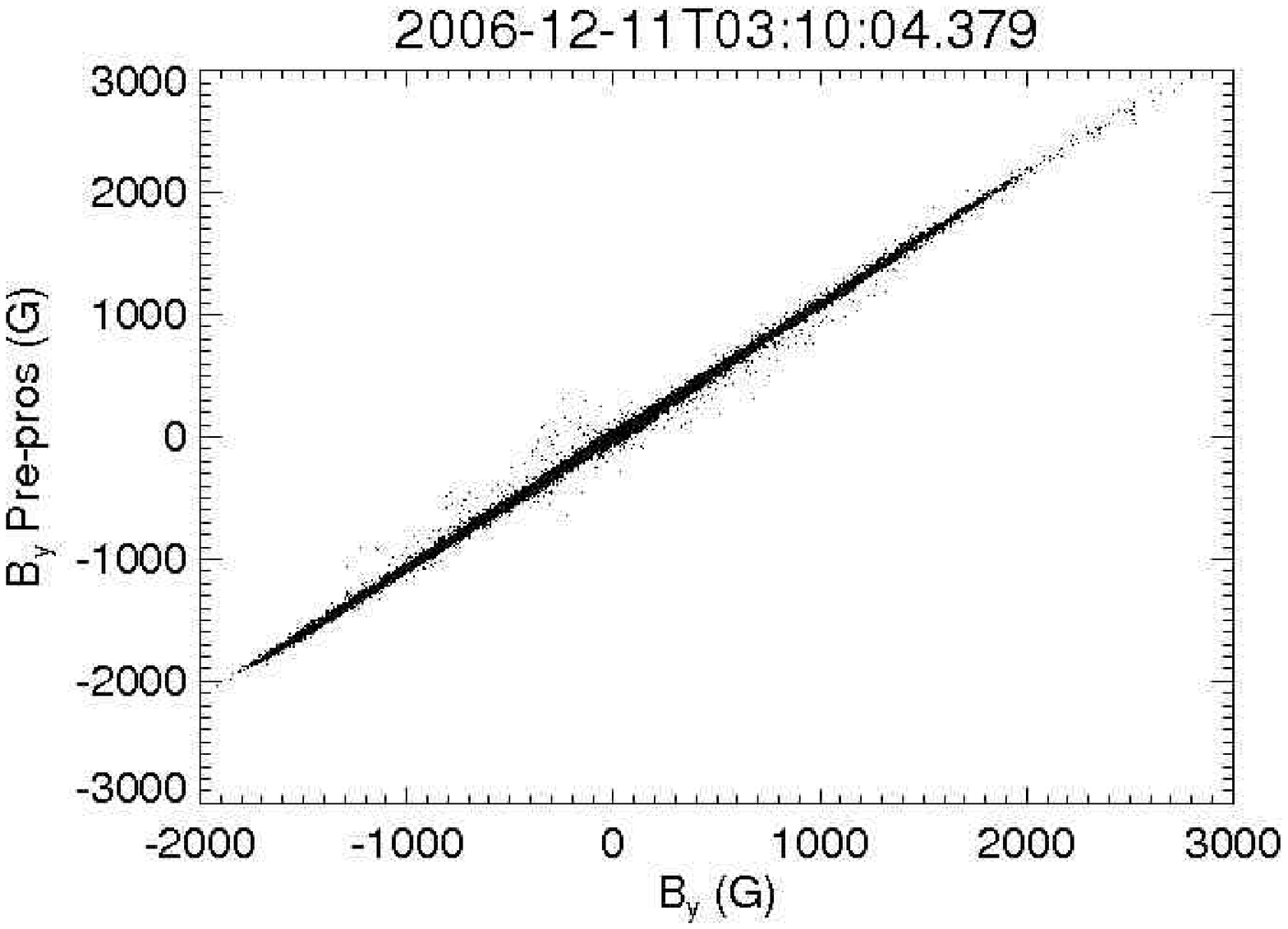} \\
\includegraphics[width=60mm]{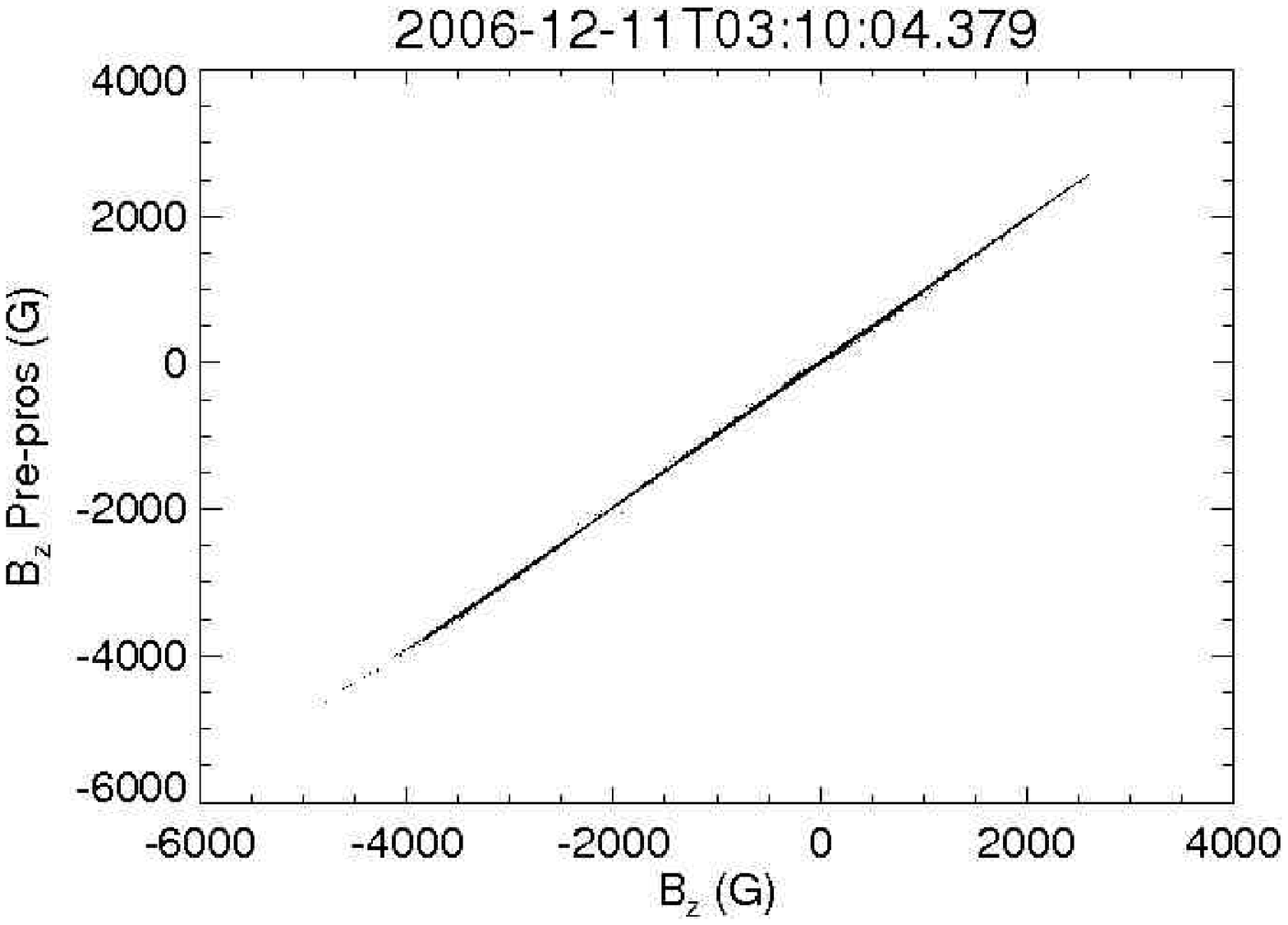} \\
\end{center}
\caption{Scatter plots of $B_{x}$(top left), $B_{y}$(top right) and $B_{z}$(bottom) 
between the photospheric and pre-processed vector field data.}
\label{fig:3}
\end{figure*}

The pre-processed vector magnetograms have been used to compute the magnetic energy.
Because of the pre-processing, the 3-components of the magnetic fields have been modified.
We compare the pre-processed vector field data with the original data. Figure \ref{fig:3}
shows the pixel-by-pixel scatter plot of the same for 3 components of the vector 
magnetic fields. We have kept the $B_{z}$ component of the magnetic field close to 
the original as suggested by Metcalf et~al. (2008). However, the pre-processing has 
changed the transverse fields significantly.  
In order to make sure that the pre-processed vector magnetograms show the 
force-freeness assumption, 
we computed the force balance parameter ($\epsilon_{force}$) and torque 
balance parameter ($\epsilon_{torque}$) as suggested in Wiegelmann, 
Inhester and   Sakurai (2006). We found that the force and torque 
balance parameters are much smaller than unity. For example, the computed 
$\epsilon_{force}$ = 0.002 and $\epsilon_{torque}$ = 0.037 for 
the December 11, pre-processed magnetogram data observed between 03:10 and 
04:19~UT. Similarly, the computed $\epsilon_{force} = -0.003$ and 
$\epsilon_{torque}$ = 0.03 for the December 11, between 08:00~UT and 09:30~UT.
This value is much smaller compared to the $\epsilon_{force}$ = 0.17 and 
$\epsilon_{torque}$ = 0.37 for the original Hinode magnetograms obtained on December 11, 
between 03:10 and 04:19 UT. A similar value is obtained for the magnetogram taken on 
December 11, between 08:00~UT and 09:30~UT.

\begin{figure*}
\begin{center}
\includegraphics[width=100mm]{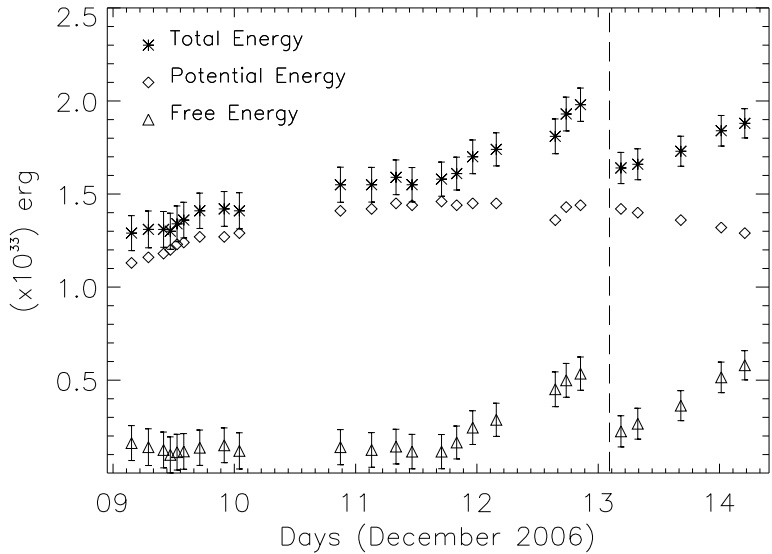} \\
\end{center}
\caption{The total magnetic energy ($\star$), the magnetic potential energy ($\diamond$) 
and the magnetic free energy ($\triangle$) available in the active region is plotted 
as a function of time. The vertical dashed line represents the onset of the X3.4~class 
flare.}
\label{fig:4}
\end{figure*}

Figure \ref{fig:4} shows the plot of the total energy ($\star$), magnetic potential 
energy ($\diamond$) and free magnetic energy ($\triangle$)
computed using Equations (1) and (2) as functions of time. The uncertainty in measuring the 
total magnetic energy involves (1) those from the measurement in the field and (2) those due to the flux imbalance term in the virial theorem. The uncertainty in the magnetic 
field measurement has been propagated
through the magnetic virial energy equation and we have estimated the uncertainty in total
magnetic energy. The term $1/8\pi\int_{\delta V}B^{2}(x.n)d^{2}x$ is neglected in the
virial energy estimation (see Metcalf et~al. 2008). This introduces an uncertainty
in the total energy estimation. We estimated the uncertainty introduced by the neglected
term in the virial energy by making the assumption that the net flux is spread
over the hemisphere uniformly. We have taken the radius of the region of interest as
$x$ in the equation and the area as the surface area of the hemisphere. The final uncertainty
is $\sqrt{error(1)^2 + error(2)^2}$, where $error(1)$ and $ error(2)$ are the uncertainties 
due to reasons (1) and (2) respectively. The uncertainties obtained from these two are
small and shown in the plot. There is one more systematic uncertainty that occurs due
to the fact that the field is not force-free in the photosphere. We have used the 
pre-processing technique to make it more closely approximating the force-free condition. However, there are
still residual net forces. These residuals produce a systematic uncertainty in the virial
energy estimation. Metcalf et~al. (2008) model shows that the net 
Lorentz force underestimates the virial energy by about 10\%. By taking this into
account our estimated total virial energy goes up by 10\% and hence also the free energy.
This systematic underestimation shifts the total energy and free energy curve 
proportionately to each other. However this does not change the difference in 
energy between before and after the flare significantly.  
From the plot it is clear that the free magnetic energy 
increases with time. However, the increase in the free energy is large on 
December 11, at 23:10~UT to December 12, at 20:30~UT, and the free 
magnetic energy accumulated over a period of 20~hrs is estimated to be about 
3.7$\times$10$^{32}$~erg. The last vector magnetogram before the X3.4 class flare was 
obtained on December 12, at 20:30~UT. The next vector magnetogram was obtained 
on December 13, at  04:30~UT. So, there is a gap of about 8~hrs between the 
magnetograms. During this period we observe a decrease in the magnetic free energy of about 
3.11$\times$10$^{32}$~erg. This drop in free energy is clearly seen in the plot after 
the X3.4 class flare (shown as the dashed vertical line for the onset time of X3.4 class
flare) has occurred. Afterwards the free energy again increases with time.

\subsection{Energy Carried Away by the Coronal Mass Ejection}
In order to compare the available magnetic free energy with the energy carried away
by the CME, we have estimated the mechanical energy in the CME.
The mass of the CME, averaged between the LASCO C2 and C3 images and assuming both the projected value and that at an angle of 
67$^{\circ}$ from the sky plane, was (8.6--9.3)$\times$10$^{15}$ g. Its speed, obtained from 
a least squares linear fit through the distance-time plot and de-projected using 
the location of the flare and Equation~(\ref{1overR}), was 1775--3060 km~s$^{-1}$. This is a fast and massive CME but not physically unreasonable. CMEs have been known to obtain 
such speeds early in their evolution but are expected to decelerate rapidly through 
the heliosphere. Given that the travel time from the Sun to the ACE spacecraft 
(judging by the arrival of a forward shock at ACE on December 14 around 13:50~UT) 
was just under $1\frac{1}{2}$ days, such a deceleration was likely. It is noteworthy that following a sudden commencement at 14:15~UT on December 14, a 
strong geomagnetic storm followed achieving a maximum K$_{\mbox p}$ of 8+ and minimum 
Dst of --146 nT. Such activity is indicative of the arrival of a large fast CME at 
the Earth.

The resulting kinetic energy of the CME was therefore (1.35--4.35)$\times$10$^{32}$~erg. 
We calculated the gravitational potential energy at each distance from the Sun but 
it never exceeded 4\%\ of the kinetic energy. Including the gravitational potential 
energy we estimate the total energy to be (1.4--4.5)$\times$10$^{32}$~erg. This is 
(0.45--1.45) times the available magnetic free energy.

\subsection{Energy Content of Thermal Plasma}

\label{sec:res}
\begin{figure*}
\begin{center}
\includegraphics[width=80mm]{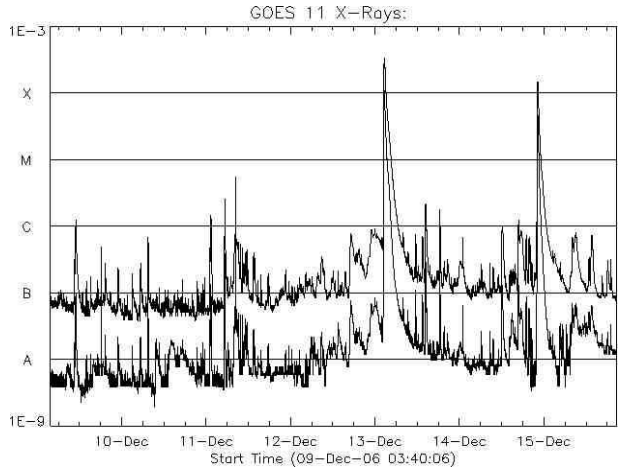} \\
\end{center}
\caption{GOES-11 X-ray flux for 5 days starting from Dec 9, 2006. The upper and 
lower curves in the plot indicates the integrated full-Sun soft X-ray flux in the 1-8 
and 0.5-4~\AA~band passes respectively. The 1-8~\AA~band pass curves give the X-ray flare 
index.}
\label{fig:5}
\end{figure*} 
Geostationary Operational Environmental Satellite (GOES) provides disk integrated 
X-ray flux in two wavelength bands, (1) 1-8~\AA~and (2) 0.5-4~\AA~(Kahler and Kreplin, 
1991). The 1-8~\AA~bandpass curves give the X-ray flare index. Figure \ref{fig:5}
shows the GOES X-ray flux obtained from GOES-11 for the two channels over 
a period of 7-days around the event of interest. 
The data from the two channels were used to estimate the radiative soft-X-ray energy. 
We first subtracted the background flux taken on December 10, 2006 between 19:09~UT and
23:12~UT. Using the solar software pipeline, the temperature of the plasma
has been estimated  by the taking the ratio of the two channels (Thomas et~al. 1985). 
Using the synthetic solar spectrum, the computed temperature can be converted into an
emission measure. 
By integrating this over the flare duration we obtained the radiated energy by the 
thermal plasma (integrated between 02:14~UT and 02:54~UT on December 13, 2006).   
The obtained radiative energy during the flare time is 9.04$\times$10$^{30}$~erg. 

It has already been shown that the ejected CME on Dec 13, 2006 carried a mechanical
energy of (1.4--4.5)$\times$10$^{32}$~erg. The radiative energy loss in the
flare plasma is 9.04$\times$10$^{30}$~erg. Adding the two shows that 
(1.49--4.59)$\times$10$^{30}$~erg of energy is utilized. In this energy estimation we do not 
consider the energy used in the production of SEPs, magnetic field reconfiguration or 
unerupted plasma transportation. 

\section{Discussion}
The magnetic energy can build up in the process of shearing motion of the magnetic
footpoints or during the flux emergence processes. The emerging flux region is the 
location for many observed solar flares and CMEs (Martin et al. 1982). The emergence 
of a magnetic field can occur on a small scale (Berlicki et~al. 2004) or large scale
(Li et~al. 2000). In the case of NOAA AR 10930, a large scale magnetic field emerged 
near the pre existing large south (negative) polarity region. From the magnetic map and 
transverse vectors it appears that the region between the 
two polarities is highly sheared (Schrijver et~al. 2008). The flux tubes emerge with 
a twist (Longcope and Welsch 2000) albeit with various quantities. While the twisted 
flux tube is emerging, it imports magnetic energy and helicity into the corona 
(Magara and Longcope 2003). In the emerging flux region the energy contribution
comes from both the horizontal motion of the footpoint as well as from the vertical
motion of emergence.  In our study, we do not know how much is the contribution 
from each of these components. This will be the subject matter of future studies. 
For the present case, however, our results establish that in an emerging active region
such as NOAA 10930 discussed here, the magnetic free energy was increasing until the onset
of X3.4 class flare and it accumulated 5.35$\times$10$^{32}$~erg of magnetic free energy
over a period of 4 days.

There have been several studies on this active region to estimate the energy released 
during the X3.4~class flare on December 13, 2006 based on NLFFF model. 
Schrijver et~al. (2008) have found
that there is a drop in energy of 3$\times$10$^{32}$~ergs during the X3.4~class flare.
Guo et~al. (2008) have obtained a 2.4$\times$10$^{31}$~ergs of drop in energy after 
X3.4~class flare. On the other hand Jing et~al. (2010) did not find any 
drop in energy, instead they observed an increase in free energy after the flare.
In the present study, we have used a magnetic virial theorem to estimate the
magnetic free energy available in the active region. Even though the cadence of the
vector field measurement is poor, we were able to see a decrease in the magnetic free
energy that persisted even 10~hrs after the flare. 

We have identified a reduction of 3.11$\times$10$^{32}$~erg of magnetic free energy 
in AR 10930. The CME associated with the energetic event that took place on 
December 13, 2006 carried about (1.4--4.5)$\times$10$^{32}$~erg (using projected 
and deprojected method) of energy. The estimated radiative 
energy loss during the X3.4 class flare was about 1$\times$10$^{31}$~ergs. The range of energies gives rise to two possibilities: (1) the entire energy for the CME came from the same active region; or (2) only a portion of the CME energy has been supplied by this active region. The latter is based on the assumption 
that the active region is associated with a single footpoint of the CME 
(Simnett and Harrison 1984, 1985, Harrison and Simnett 1984, Harrison 1986). If the calculated energy is closer to the projected calculations (the lower limit) then the former is likely true, while if the energy is closer to the de-projected values then the latter applies.

The latter possibility is more likely for this event given that it is a halo CME and therefore likely to be highly projected in the LASCO images. Let us consider the upper extreme of the energy. Half of the CME energy is around 
2.25$\times$10$^{32}$~erg, which leaves 8.6$\times$10$^{31}$~erg in 
the lost magnetic free energy available for other purposes.
Emslie et~al. (2004) determined the energy budget on two flare/CME events on 
21~April and 23~July 2002. Both were associated with an X-class flare (X1.5 for 
the former and X4.8 for the latter). They found that the flares contributed between 
1-4$\times$10$^{31}$~erg, which included the energy for particle acceleration. Given 
that flares are broadband phenomena the emission energy contribution extends across 
the electromagnetic spectrum, and so a larger energy budget for emission is expected.
We must also consider the energy in moving around solar plasma and un-erupted
magnetic fields (e.g. Webb et~al. 1980). It is not unreasonable to conclude that 
this energy budget combined could accommodate the remaining magnetic energy 
($\sim$8.6$\times$10$^{31}$~erg) that did not contribute to the CME. 

At this stage, however, we do not have sufficient evidence to determine which possibility is correct, but it is noteworthy that the calculated energy range (for the CME) includes the 
value for the available free energy, which lies almost exactly halfway (by ratio) between 
the two extremes.

The actual magnetic free energy release 
could also be larger than that computed. This is because of the following reasons:
\begin{enumerate} 
\item The magnetic flux was still emerging before the flare (Park et~al. 2010) 
and our preflare magnetic
field measurement was about 5~hrs before the flare. So, the available free energy
could be larger than estimated here. 
\item The post-flare energy calculation was performed at a time one hour after the 
post-flare time. Within this 1~hr gap the free energy would have increased as the 
flux was still increasing (Park et~al. 2010). Hence the measured dip in free energy 
could be larger. 
\item The photosphere is not force-free. To make it close to 
the force-free  we have used the pre-processing technique that may reduce the 
actual value of the available energy. Metcalf et~al. (2008) have applied a realistic
model to test the Lorentz forces in the photosphere. In their model they compared 
the net force at various heights. They have found that the net force falls close to 
zero at the chromospheric height but not in the photosphere. Due to this net force
at the photospheric level there is an underestimation of the virial energy by about 
10\%. Hence, the free energy will also be underestimated by the same amount. However, 
this underestimation will not affect significantly the change in free magnetic 
energy measurement between before and after the flare. 
\item It is not well known how much the 180$^{\circ}$ ambiguity will affect the 
virial energy estimation. 
\end{enumerate}

We have identified the source of uncertainty in the energy estimation using the virial theorem.
The estimated uncertainty is much smaller than the free energy. Metcalf, Leka and Mickey (2005) 
have estimated the uncertainty in magnetic energy estimation by pseudo-Monte Carlo method
by displacing the origin to different locations on the vector fields. This 
method is valid only if the measured field is forced. However, the pre-processing technique makes these fields close to force-free and hence measuring
the mean value of the energy by shifting the origin to different locations on the image
is not required. Wheatland and Metcalf (2006) have introduced a new method to find the magnetic
energy using the virial theorem. However, that method depends on the choice of the model
to compute the vertical field gradients. In the absence of chromospheric magnetic field
measurements they have used the linear force-free parameter ($\alpha$) to compute the
vertical field gradients. The distribution of magnetic field deviates from the 
constant $\alpha$ in most of the locations in the active region. Hence, a detailed study
is required to find further details on how this method is superior to the other methods.
Until then the method of using the pre-processed data for the virial energy estimation 
is a better choice than those methods.  

The largest source of error in the CME calculations is the 3-D information provided for
the de-projection. We used the location of the flare to provide values for $\theta$, 
$\phi$ (distance) and $\chi$ (mass) but this assumption is limited. Firstly, it is not
appropriate to assume that the CME originated from the flare or even the active region, 
and secondly the CME cannot be regarded as a point source (as assumed for the distance 
measurements). The CME is a large structure and for the larger ones the associated flare
and active region are typically associated with a single footpoint only. In this case, 3-D 
assumptions based on the flare and AR would provide information on just that footpoint. 
This is also the reason why we assume for the maximum case that only half of the CME's energy was 
provided by the AR. To our knowledge no systematic empirical study of the source of energy for 
the CME has yet been conducted so it is unknown what proportion of its energy budget 
arises from the AR. Given the values we arrived at from the present study, however 
limited, it seems that a 50\%\ contribution may be a reasonable first-order 
assumption. Such knowledge may lead to a greater understanding of the onset mechanism 
for CMEs.

Jing et~al. (2009) have observed for many active regions that there is a decrease
in magnetic free energy 15~min before the peak time of the associated non-thermal 
flare emission. Such studies cannot be taken up with 
the present data as the cadence of vector magnetic field measurement is poor.
The Solar Dynamic Observatory (SDO) (launched February 2010) provides vector 
field measurements at high cadence. Hence, the SDO data set may provide a complete 
picture on how the free energy changes close to the flare time and during the flare.
However, for a better estimation of the available free energy we need to have chromospheric
vector magnetic field data (Metcalf, et~al. 2005). Currently, the chromospheric vector
field data are not available on regular basis. Until they are available, photospheric 
vector magnetogram with the pre-processing techniques will provide the good 
boundary data for estimating the magnetic free energy. 
 
\section{Conclusions}
Solar flares and CMEs are responsible for the release of 
large amounts of energy from the Sun. During solar flares energy is redistributed
partly in the form of broadband emission from X-rays to the radio band. The
energy is also carried by the thermal plasma, accelerated electrons, etc.
The vast majority of the energy budget, however, is allocated to mechanical energy 
in the form of an expanding CME. 

Using the high resolution time sequence of vector magnetograms from SOT/SP we studied the 
temporal evolution of magnetic free energy in an active region
NOAA 10930.  We then compared the available magnetic free energy with the energy
carried away by the halo CME.
From the analysis we arrive at the following conclusions:
\begin{enumerate}
\item  The free magnetic energy increases with time 
and about 3.7$\times$10$^{32}$~erg of magnetic free energy is accumulated
over a period of about 20~hrs before the X3.4 class flare and CME launch occurred. 
\item There is a decrease in magnetic free energy of about 3.11$\times$10$^{32}$~erg during
the flare/CME. 
\item The estimated energy carried away by the CME is (1.4--4.5)$_\times$10$^{32}$~ergs (using 
projected and de-projected values). This
is 0.5--1.5 times larger than the estimated magnetic free energy.
\item The estimated radiative energy loss during the X3.4 class flare is 
9.04$\times$10$^{30}$~erg.

We believe that more energy was carried by the CME than was available in the active region, 
as some of the free energy must be allocated to radiative energy loss, particle acceleration 
and plasma and magnetic field reorientation. Given the range of energies calculated for CME and the possibility that the free magnetic energy may be larger, one could easily conclude that the entire energy was provided by the active region.
\end{enumerate}

\section*{Acknowledgments}
We would like to thank the anonymous referee for his/her constructive comments
which improved the clarity of the manuscript.
Hinode is a Japanese mission developed and launched by
ISAS/JAXA, with NAOJ as domestic partner and NASA and
STFC (UK) as international partners. It is operated by these
agencies in co-operation with ESA and the NSC (Norway).
SOHO is a project of international cooperation between
ESA and NASA. TH's work is supported in part by the NSF SHINE 
competition, Award 0849916.

\label{lastpage}

\begin{thebibliography}{99}
\bibitem{b1}  Alissandrakis, C. E. 1981, A\&A, 100, 197
\bibitem{b2} Aly, J. J. 1984, ApJ, 283, 349.
\bibitem{b3} Berlicki, A., Schmieder, B., Vilmer, N., Aulanier, G. and 
Del Zanna, G., 2004, A\&A 423, 1119.
\bibitem{b4} Billings, D.E., 1966, A Guide to the Solar Corona, Academic Press,
New York. 
\bibitem{b5} Brueckner, G.~E., Howard, R.~A., Koomen, M.~J., Korendyke, C.~M., 
Michels, D.~J., Moses, J.~D., Socker, D.~G., Dere, K.~P., Lamy, P.~L.,
Llebaria, A., Bout, M.~V., Schwenn, R., Simnett, G.~M., Bedford, D.~K., 
and Eyles, C.~J., 1995, Solar Phys., 162, 357.
\bibitem{b6} Canfield, R.~C., Cheng, C.-C., Dere, K.~P., Dulk, G.~A., McLean, D.~J.,
Robinson, Jr., R.~D., Schmahl, E.~J., and Schoolman, S.~A., 1980, in Solar Flares: 
A Monograph From Skylab Workshop II, Sturrock, P.~A (ed.), Colo. Assoc. Uni. Press, 
Boulder Colo, 451.
\bibitem{b7} Chandrasekhar, S. 1961, Hydrodynamic and Hydromagnetic Stability
(NewYork:Dover).
\bibitem{b8} De Rosa, M. L., Schrijver, C. J.; Barnes, G. Leka, K. D., Lites, B. W. 
and Aschwanden, M. J., et~al. 2009, ApJ, 696, 1780.
\bibitem{b9} Emslie, A. G., Kucharek, H., Dennis, B. R., Gopalswamy, N., 
Holman, G. D., Share, G. H., et~al. 2004, JGR, 109, A10104.
\bibitem{b10} Gary, G. A., Moore, R. L., Hagyard, M. J., and Haisch, B. M., 1987, 314,
782.
\bibitem{b11} Gary, G. A., 1989, ApJS, 69, 323.
\bibitem{b12} Guo, Y. Ding, M. D., Wiegelmann, T., and Li H., 2008, ApJ, 679, 1629.
\bibitem{b13} Harrison, R.~A., 1986, A\&A, 162, 283.
\bibitem{b14} Harrison, R.~A., Simnett, G.M., 1984, Adv. Space Res. 4, 199.
\bibitem{b15} Howard, T.~A., Fry, C.~D., Johnston, J.~C., and Webb, D.~F., 2007,
ApJ, 667, 610.
\bibitem{b16} Howard, T.~A., Nandy, D., and Koepke, A.~C., 2008, JGR, A01104,
doi:10.1029/2007JA012500.
\bibitem{b17} Howard, T.~A., and Tappin, S.~J., 2008, Solar Phys., 252, 373.
\bibitem{b18} Houminer, Z., and Hewish, A., 1972, Planet. Space Sci., 20, 1073.
\bibitem{b19} Ichimoto, K., Lites, B., Elmore, D., Suematsu, Y., Tsuneta, S., et~al.
2008, Solar Phys. 249, 233.
\bibitem{b20} Isobe, H., et al. 2007, PASJ, 59, 807
\bibitem{b21} Jing, Ju, Chen, P. F., Wiegelmann, T., Xu, Y., Park, S. H., 
and Wang, H., 2009, ApJ 696, 84.
\bibitem{b22} Jing, Ju, Yuan, Y. Wang, B., Wiegelmann, T., Xu, Y. and Wang, H.
2010, ApJ, 713, 440.
\bibitem{b22} Kahler, S. W., and Kreplin, R. W., 1991, SoPh, 133, 371.
\bibitem{b23} Kazachenko, M. D., Canfield, R. C., Longcope, D. W., Qiu, J.,
Des Jardins, A. and Nightingale, R. W., 2009, ApJ, 704, 1146.
\bibitem{b24} Kosugi, T., Matsuzaki, K., Sakao, T. et~al. 2007, Solar Phys.,
243, 3.
\bibitem{b25} Leka, K. D., Barnes, G.,  and Crouch, A. D. 2009, In Second Hinode
Science Meeting, ASP Conference Series, 416, 127.
\bibitem{b26} Li, H., Sakurai, T. Ichimoto, K. and UeNo, S., 2000, PASJ, 52, 483.
\bibitem{b27} Lin, H., Kuhn, J. R. and Coulter, R., 2004, ApJ, 613, 177.
\bibitem{b28} Lites, B. W. and Skumanich, A. 1990, ApJ, 348, 747.
\bibitem{b29} Lites, B. W., Elmore, D. F., Seagraves, P., and Skumanich, A. P., 
1993, ApJ, 418, 928.
\bibitem{b30} Longcope, D. W. and Welsch, B. T., 2000, ApJ, 545, 1089.
\bibitem{b31} Longcope, D. W.,  Beveridge, C., Qiu, J., Ravindra, B., Barnes, G., 
and Dasso, S., 2007, Solar Phys., 244, 45.
\bibitem{b32} Low, B. C., 1985 in Measurement of Solar Vector Magnetic Fields, ed. 
M. J. Hagyard (NASA Conf. Pub. 2374), 49.
\bibitem{b33} Magara, T. and Longcope, D. W., 2003, ApJ, 586, 630.
\bibitem{b34} Magara, T., and Tsuneta, S. 2008, PASJ, 60, 1181.
\bibitem{b35} Martin, S. F., Dezso, L., Antalova, A., Kucera, A. and Harvey, K. L.
1982, AdSR, 2, 39.
\bibitem{b36} Metcalf, T., 1994, Solar Phys., 155, 235.
\bibitem{b37} Metcalf, T. R., Jiao, L, McClymont, A, N.; Canfield, R, C. and 
Uitenbroek, H., 1995, ApJ, 439, 474.
\bibitem{b38} Metcalf, T. R., Leka, K. D. and Mickey, D. L., 2005, ApJ, 623, 53.
\bibitem{b39} Metcalf, T. R., De Rosa, M. L., Schrijver, C. J., Barnes, G. et~al.,
2008, Solar Phys., 247, 269.
\bibitem{b40} Min, S. and Chae, J., 2009, Solar Phys., 258, 203.
\bibitem{b41} Molodenskii, M. M., 1969, SVA, 12, 585.
\bibitem{b42} Park, S.~H., Chae, J., Jing, Ju, Tan, C., and Wang, H., 2010, ApJ, 720, 1102.
\bibitem{b43} Ravindra, B., Longcope, D. W. and Abbett, W. P., 2008, ApJ, 677, 751.
\bibitem{b44} Sakurai, T., 1987, Solar Phys., 113, 137.
\bibitem{b45} Schrijver, C. J., De Rosa, M. L., Metcalf, T., Barnes, G., Lites, B.,
Tarbell, T., McTiernan, J., Valori, G., Wiegelmann, T., Wheatland, M. S., et~al.,
2008, ApJ, 675, 1637.
\bibitem{b46} Skumanich, A., and Lites, B. W. 1987, ApJ, 322, 473.
\bibitem{b47} Simnett, G.~M., Harrison, R.~A., 1984, Adv. Space Res., 4, 279.
\bibitem{b48} Simnett, G.~M., Harrison, R.~A., 1985, Solar Phys. 99, 291.
\bibitem{b49}  Solanki, S. K., Lagg, A., Woch, J., Krupp, N. and Collados, M., 2003,
Nature, 425, 692.
\bibitem{b50} Srivastava, N., Mathew, S. K., Rohan, L. E., Wiegelmann, T., 2009, JGRA,
11403107.
\bibitem{b51} Su, J. T., Sakurai, T., Suematsu, Y., Hagino, M. and Liu, Yu, 2009, 
ApJ, 697, 103.
\bibitem{b52} Tan, C., Chen, P. F., Abramenko, V., and Wang, H. 2009, ApJ, 690, 1820.
\bibitem{b53} Thosmas, R.~J., Crannell, C.~J., and Starr, R., 1985, Solar Phys. 95, 323.
\bibitem{b54} Venkatakrishnan, P., Hagyard, M. J., and Hathaway, D. H. 1988, 
Solar Phys., 115, 125.
\bibitem{b55} Venkatakrishnan, P and Ravindra, B., 2003, GRL, 30, 2181.
\bibitem{b56} Webb, D.~F., Cheng, C.-C., Dulk, G.~A., Edberg, S.~J., Martin, S.~F.,
McKenna-Lawlor, S., and McLean, D.~J., 1980, in Solar Flares: A Monograph From Skylab Workshop
II, Sturrock, P.~A (ed.), Colo. Assoc. Uni. Press, Boulder Colo, 471.
\bibitem{b57} Wheatland, M. S., and Metcalf, T. R., 2006, ApJ, 636, 1151.
\bibitem{b58} Wiegelmann, T., Inhester, B., and Sakurai, T., 2006, Solar Phys.,
233, 215.
\bibitem{b59} Wiegelmann, T., Thalmann, J. K., Schrijver, C. J., De Rosa, M. L.,
and  Metcalf, T. R., 2008, Solar Phys., 247, 249.
\bibitem{b60} Yashiro, S., Gopalswamy, N., Michalek, G., St.~Cyr, O.~C., Plunckett S.~P.,
and Howard, R.~A., 2004, JGR, 109, doi:10.1029/2003JA010282.

\end{thebibliography}
\end{document}